\documentclass[aps,pra,showpacs,twocolumn,superscriptaddress,longbibliography]{revtex4-1}
\usepackage{graphicx}
\usepackage[usenames]{color}
\usepackage{amssymb,amsmath}
\usepackage{siunitx}
\usepackage{xcolor}
\usepackage{setspace} 
\usepackage{float} 
\usepackage{tabularx}

\newcommand{\fig}[1]{Fig.~\ref{#1}}
\newcommand{\be}[1]{\begin{equation}\label{#1}}
\newcommand{\ee}{\end{equation}}

\usepackage[LGRgreek]{mathastext}
\usepackage{lipsum}

\begin{document}

\title{Triple ionization and ``frustrated" triple ionization in triatomic molecules driven by intense laser fields }

\author{M. B. Peters }
\affiliation{Department of Physics and Astronomy, University College London, Gower Street, London WC1E 6BT, United Kingdom}
\author{V. P. Majety}
\affiliation{Department of Physics and Center for Atomic, Molecular and Optical Sciences and Technology, Indian Institute of Technology Tirupati, Tirupati 517506, India }
\author{A. Emmanouilidou}
\affiliation{Department of Physics and Astronomy, University College London, Gower Street, London WC1E 6BT, United Kingdom}

\begin{abstract}
We formulate a three-dimensional semi-classical model to treat three-electron escape dynamics in a strongly-driven linear  triatomic molecule, HeH$_{2}^{+}$. Our model includes the Coulomb singularities. Hence, to avoid unphysical autoionization, we employ two criteria to switch off the Coulomb repulsive force between two bound electrons and switch it on when the motion of one 
electron is mostly determined by the laser field. We investigate triple and ``frustrated" triple ionization. In the latter process two electrons escape while one electron remains bound in a Rydberg state.  We find that two pathways prevail in ``frustrated" triple ionization, as in ``frustrated" double ionization. We also find that the electron that remains in a Rydberg state is more likely to be attached to He$^{2+}$ compared to H$^{+}$. Our results indicate that in triple and ``frustrated" triple ionization  electronic correlation is weak. Moreover, we compute the sum of the kinetic energies as well as the angular patterns of the final ion fragments in triple and ``frustrated" triple ionization. These patterns suggest that the fragmenting molecule deviates from its initial linear configuration.

 \end{abstract}
\date{\today}

\maketitle
\section{Introduction}
Correlated multi-electron escape dynamics arising in systems driven by intense infrared and mid-infrared laser fields is a problem of fundamental interest.
The complexity of the problem, currently, limits  ab-initio quantum mechanical computations to two-electron escape in strongly-driven atoms  \cite{abinitio1,abinitio2}. This latter problem has also been addressed by  three-dimensional quantum mechanical  \cite{Scrinzi} and semiclassical techniques \cite{semi-classical3, semi-classical4} that include the Coulomb singularity. Given the even larger degree of complexity, strongly-driven three-electron dynamics 
 has been addressed in few theoretical  \cite{TI1,TI2,TI3,TI4} and experimental studies \cite{exp31,exp32}. More relevant to the current work is the classical study of driven trimers in ref. \cite{TI4}  with three atoms placed far apart and each electron being bound to a different atom. The work in  ref. \cite{TI4} does not address the unphysical autoionization that occurs in a classical treatment of two bound electrons when the Coulomb singularities are included. One electron can acquire a very negative energy and release energy that leads to the escape of the other bound electron. This does not occur quantum mechanically, since the energy of an electron has a lower bound. This unphysical autoionization is addressed in this semiclassical work that includes the Coulomb singularities and involves two bound electrons.

 Here, we develop a three-dimensional (3D) semiclassical model to investigate three electron dynamics in strongly-driven triatomic molecules. We do so in the context of the strongly-driven linear molecule HeH$_{2}^{+}$. This model is an extension to the 6-body Coulomb problem of the model we developed to describe, first, H$_{2}$  \cite{Mol1,Mol2}, and,  then,  D$_{3}^{+}$\cite{Mol3} when  driven by intense laser fields.  Treating unphysical autoionization is an aspect of our model introduced in this work. The latter arises since we fully account for the Coulomb singularities and in 
 HeH$_{2}^{+}$  more than one electron can be bound.

Using  this 3D semiclassical model, we account  for triple and double  ionization as well as  for ``frustrated" triple  ionization. 
``Frustrated" ionization involves the formation of Rydberg states. Namely, 
 an electron first tunnel ionizes in the driving laser field. Then, due to the electric field, this electron is recaptured by the parent ion in a Rydberg state \cite{Nubbemeyer}. In ``frustrated" double ionization (FDI) an electron is ionized while another one remains bound in a Rydberg state at the end of the laser pulse. 
 ``Frustrated" double ionization  accounts for roughly 10\% of all ionization events. Hence, FDI is  a major process  in the breakup of  strongly-driven molecules. It has been addressed in  experimental studies   of H$_{2}$ \cite{Manschwetus}, D$_{2}$ \cite{Zhang2} and of the two-electron triatomic molecules D$_{3}^{+}$ and H$_{3}^{+}$ \cite{McKenna1, MScKenna2, Sayler}.   
 
 Two pathways account for ``frustrated" double ionization in strongly-driven two-electron diatomic and triatomic molecules \cite{Mol1,Mol2,Mol3}. In both pathways, one electron tunnel ionizes early on (first step), while  the remaining bound electron  ionizes later in time (second step). In pathway A it is the second step that is frustrated, i.e. the initially bound electron does not escape but remains bound in a Rydberg state \cite{Mol1}. In pathway B it is the first step that is frustrated, i.e. the initially tunneling electron is captured in a Rydberg state \cite{Mol2}.  Also, electron-electron correlation, which can underlie pathway B  \cite{Mol1, Mol3}, can be controlled with orthogonally polarised two-color linear laser fields  \cite{Mol6,CTP}. Furthermore, significant  enhancement of pathway B of FDI with no electronic correlation  is achieved when driving  triatomic molecules with counter-rotating two-color circular laser fields      \cite{Mol5}. It was shown that this is due to the electron that tunnel-ionizes first ``hovering" around the nuclei. This feature is most prevalent when 800 nm and 400 nm laser fields are employed with a field-strength ratio equal to two \cite{Mol5}. In addition, 
  fingerprints of nuclear motion on the electron dynamics  have been previously identified in ``frustrated" double ionization \cite{Mol4,Zhang}. Such a signature includes an oscillation in the principal n quantum number \cite{Mol4}.

In this work, we address ``frustrated" triple ionization (FTI) where two electrons escape while one remains bound in a Rydberg state. We identify the pathways of ``frustrated" triple ionization and compute the principle n quantum number in FTI. Moreover, we compute triple as well as double ionization and discuss the role that correlation plays in the three- and  two-electron escape. We also compute the distributions of the kinetic energy release and of the angles of the final ion fragments in all three ionization processes.

 \section{Method}
 In the initial state  of HeH$_{2}^{+}$,  all three atoms are placed along the z-axis. The two hydrogen atoms are at -3.09 a.u. and  -1.02 a.u., respectively, and the helium atom is at  1.04 a.u. We refer to H farther away from He as left H and the one closest to He as middle H. We compute the distance between the two hydrogen atoms and the hydrogen and helium atoms using the quantum chemistry package MOLPRO \cite{Molpro}. We employ the Hartree-Fock method with the aug-cc-pV5Z  basis set. The Hartree-Fock method overestimates by a small amount the distance between the hydrogen and the helium atoms \cite{initialstate}. However, we employ this method for consistency with the Hartree-Fock wavefunctions that we use
 in the potential energy terms involved in computing the exit point of the tunnel-ionizing electron \cite{Mol2}.  The electric  field is along the axis of the linear molecule, with a strength  within the below-the-barrier ionization regime.  As a result, one electron (electron 1) tunnel ionizes at time t$_{0}$
 through the field-lowered Coulomb potential. This is a quantum-mechanical step. We employ a quantum-mechanical calculation to compute this ionization rate. Specifically, we obtain the alignment-dependent tunnel-ionization rate for HeH$_{2}^{+}$ by employing the hybrid anti-symmetrized coupled channels (haCC) method described in ref. \cite{Majety,Majety1}. In haCC, the system is represented in a basis of neutral and single ionization channel functions. The ground state of HeH$_{2}^{+}$ and the energetically lowest few HeH$_{2}^{+}$ states are obtained from the quantum chemistry package COLUMBUS \cite{columbus}. A purely numerical basis is used to represent the tunneling electron, while anti-symmetrization is fully enforced. Exterior complex scaling is employed in order to obtain tunnel-ionization rates. We assume that electron 1 exits along the direction of the laser field \cite{Mol2}.  We compute the first ionization energy of HeH$_{2}^{+}$ with MOLPRO
 and find it equal to 1.02 a.u.  When the tunnel electron  exits the field-lowered Coulomb barrier, we consider the momentum of the electron parallel to the field to be equal to zero. The  transverse momentum is given by a Gaussian distribution. The latter arises from standard tunneling theory  \cite{tunn1,tunn2,tunn3} and   represents the Gaussian-shaped filter with an intensity-dependent width. The initially two bound electrons (electrons 2 and 3), are each represented by a microcanonical distribution for a triatomic molecule \cite{Chen}. Each electron is assigned an energy equal to 2.21 a.u., which is half the ground state energy of HeH$_{2}^{2+}$. Hence, in the initial state, electronic correlation is only indirectly taken into account  via the energies considered in the microcanonical distributions. We initialise the nuclei at rest. Our studies suggest that an initial predissociation does not significantly alter the ionization dynamics \cite{Mol2}.

 We use an electric   field of the form

 \begin{align}
  \vec{\mathcal{E}}(t) =& \mathcal{E}_0  \exp\left[ -2\ln 2 \left(\frac{t}{\tau}\right)^2 \right]    \cos \omega_1 \mathrm{t} \hspace{0.1cm}\mathrm{\hat{z}} ,
  \label{eqnew}
 \end{align}
where $\tau=40$ fs is the full width at half maximum of the pulse duration in intensity. The electric field strength $\mathcal{E}_0$ is taken equal to 0.08 a.u. We find that the threshold of the field strength for over-the-barrier ionization is  equal to 0.087 a.u. 
 We first select randomly  the tunnel-ionisation time $t_{0}$   in the  time interval $[-2\tau,2\tau]$ and we specify   
 the initial conditions. Then,  the position and momentum of the three electrons and the three nuclei are propagated classically. We do so by employing the Hamiltonian for the 6-body Coulomb problem when driven by an intense laser field.  We account fully with no approximation for the Coulomb forces and the interaction of each electron and  nucleus with the laser field. In addition, the Coulomb singularities are explicitly included in our model  \cite{Mol2}. Moreover, the electron and nuclear dynamics are  treated on an equal footing. We employ the Wentzel-Kramers-Brillouin approximation to allow each electron to tunnel with a quantum-mechanical probability during time propagation  \cite{Mol1, Mol2,Mol3, Mol4}.  This description ensures that we accurately compute enhanced ionization   \cite{Niikura,EI1,EI2,EI3,EI4}. Regarding enhanced ionization, when the nuclei are at a critical distance, a double-potential well is formed such that it is easier for an electron bound to the higher potential well to tunnel to the lower potential well  and then  ionize.  A very good  agreement of our previous results for  H$_{2}$ \cite{Mol1} and D$_{3}^{+}$ \cite{Mol3} with experimental results \cite{Manschwetus, MScKenna2} justifies the approximations we consider in both the initial state and the time propagation 
in our model.

Here, we extend our  model to treat unphysical autoionization between two bound electrons. We do so by introducing criteria to switch  off the Coulomb repulsion between two bound electrons. Switching off the Coulomb repulsion  has been implemented in previous work 
on classical calculations of electron impact  on two-electron targets \cite{Olson}. Here, in the initial state, we switch off the correlation between the bound electrons 2 and 3. Moreover, the Coulomb repulsion is  switched on between electron 1, which tunnel-ionizes in the initial state, and the other two bound electrons.  During  time propagation,  electron 2 or 3 can  ``quasi-ionize" or ionize. When this happens,  we turn on the Coulomb repulsion between this newly ``quasi-ionized" or ionized electron and any remaining bound electrons. Moreover, if electron 1 becomes bound during time propagation,  we turn off the Coulomb repulsion between electron 1 and any other bound electrons. 

Hence, we must   identify  the time an electron ionizes or ``quasi-ionizes". We first define the time of ionization. In our previous work \cite{semi-classical4}, the ionization time, $t_{i}^{c,1}$, 
of an electron i was defined as the time when the compensated energy $(p_{x,i}^{2}+p_{y,i}^2+(p_{z,i}-\mathcal{A}(t))^2)/2-Z/r_{i})$ becomes positive and remains positive thereafter  \cite{Leopold};$\mathcal{A}(\mathrm{t})$ is the vector potential. However, in our previous studies of strongly-driven two-electron atoms 
and molecules the Coulomb force between the two electrons was switched on at all times.  Hence,  there was no need to compute the  ionization time of each of the two electrons  on the fly, i.e. during  time propagation. This time was computed only after we have registered the events corresponding to different ionization processes, during the analysis of the trajectories.
Here, for three electrons, we keep the Coulomb force between two bound electrons switched off to avoid unphysical autoionization. We employ the ionization time which is computed on the fly in order to determine when to switch on or off the Coulomb repulsion between two electrons. This time is not   $t_{i}^{c,1}$, since  the latter can not be determined during propagation. Instead, we employ the time $t_{i}^{c,2}$  when the compensated energy of electron i converges to a positive value, with $t_{i}^{c,2}>t_{i}^{c,1}$.

However,  an electron can transfer energy to the other electrons  while it is not ionized. Indeed, the compensated energy of electron 1
 is negative during its exit from the barrier and up until  electron 1 first returns to the nuclei to transfer energy to the other electrons. In our computation, this is the case for most trajectories. During this time, we find that the motion of electron 1 along the z-axis is mostly influenced by the field. That is, in half the laser period  T, the position of electron 1 along the z-axis has no more than one maximum. Hence,  we monitor the time interval between two subsequent maxima in the position of each electron along the z-axis. If there is  less than one maximum in a time interval T/2, we register as   ``quasi-ionization" time $t_{i}^{q}$ the time at the  end of this interval. Similarly, $t_{i}^{q}$ can be identified from the minima of the position of the electron along the z-axis.

We use the above two criteria to find when an electron is ionized, ``quasi-ionized" or  bound. In our computations, for roughly 90\% of FTI and triple and double ionization events, we find that the Coulomb repulsion between electrons 1 and 2 (V$_{12}$) and between electrons 2 and 3 (V$_{13}$) is never switched off. For the rest of the events, these forces  are   switched off  and on only once. Moreover, the Coulomb repulsion between the initially bound electrons 2 and 3 (V$_{23}$) is  switched on and remains on only once for almost all ionization events. This switch  on roughly takes place around five periods of the laser field after the start of the time propagation.

It follows from the above that the transfer of energy from electron 1 to the other two initially bound electrons is accounted for very well. Indeed, for most events V$_{12}$ and $V_{13}$ are never turned off. If electron 2 or 3 becomes ``quasi-ionized" or ionized during propagation, the transfer of energy from that electron to another electron is accounted for with a delay. Namely, we find that for the majority of ionization events  $t_{i}^{c,2}$ or $t_{i}^{q}$ is  larger than $t_{i}^{c,1}$ by at most T/2. Hence, the V$_{23}$ potential is switched on with a delay.  As a result, processes that are not well accounted for involve   two steps. First, they involve a transfer of energy from one electron to a bound one. Then, within a very small time interval, they involve  a second transfer of energy  from the target electron in the first transfer of energy   to another bound electron. However, we do not expect such processes to play a significant role  in strongly-driven HeH$_{2}^{+}$. In contrast, in  triple ionization by single-photon absorption  the two major pathways of ionization involve the  transfer of energy via  a sequence of two collisions that are only a few attoseconds apart. In one of the two major pathways, the target electron in the first transfer of energy, i.e. collision,  becomes the impacting one in the second collision  \cite{Agapi1,Agapi3}.

\begin{figure*}
\includegraphics[scale=0.4]{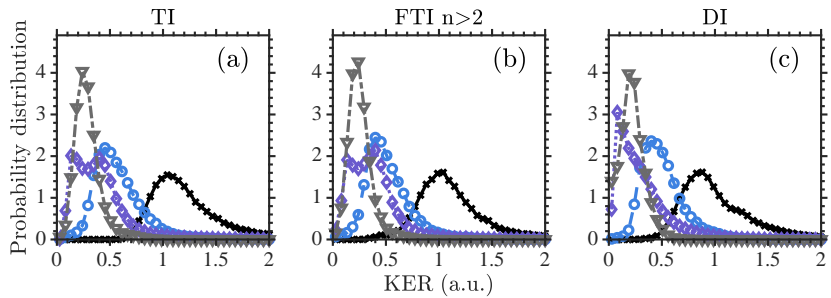}
\centering
\caption{Distribution of the sum of the final kinetic energies (black solid lines with crosses) of the ion fragments produced in (a) triple ionization, (b) ``frustrated" triple ionization and (c) double ionization. The  dashed light grey lines and downwards triangles depict the distribution of the final kinetic energy of the He$^{2+}$ ion fragment for TI,  He$^{+*}$  for FTI and He$^{+}$ for DI. The dashed purple lines and diamonds (dashed blue lines and circles) depict the distribution of the final kinetic energy of the middle (left) H$^{+}$ ion fragment for TI,  FTI and  DI. All curves are normalized to one.}
\label{fig1}
\end{figure*}

  \section{Results}
Using the 3D model described above, we focus on  triple ionization, ``frustrated" triple ionization and double ionization. We find that out of all ionization events roughly 3.5\% are TI events, 1\% are FTI events while 25\% are DI events.  
 In triple ionization the resulting fragments are He$^{2+}$ and two H$^{+}$ ions.  In ``frustrated" triple ionization,
 one electron stays in a Rydberg state either on H$^{+}$ or He$^{2+}$. However, for FTI, we find that  the  formation of He$^{+*}$ and two H$^{_+}$ ions is three times more likely than the formation of He$^{2+}$, H$^{*}$ and H$^{+}$.  Hence, in what follows we focus on the most probable channel of FTI. Moreover, we  consider FTI events in high Rydberg states with $n>2$.   The reason we ignore FTI events with $n=2$ is the same as for our work on HeH$^{+}$
  \cite{Mol4}. Namely, an electron from the $n=1$ state of H$^{+}$ tunnels to the $n=2$ state of He$^{2+}$. As a result, we obtain a large number n=2 states. 
   For DI the electron that does not ionize remains bound mostly in the $n=1$ state. Moreover, for DI it is significantly more likely for the final fragments to be He$^{+}$ and two H$^{+}$ ions rather than He$^{2+}$, H$^{+}$ and H. Hence, in what follows 
 we focus on the most probable channel  of DI as we do for  FTI, unless we  indicate otherwise.   
 
 \subsection{KER distributions}
 In \fig{fig1}, we plot the kinetic energy release (KER) distributions of the final ion fragments for triple ionization, ``frustrated" triple ionization and double ionization. We find that the KER distribution  peaks around 1 a.u. for TI and FTI, while it peaks around 0.8 a.u. for DI. These peak values are consistent with the peak values of the distributions of the inter-nuclear distances at the time an electron  tunnel-ionizes last. Indeed, we find (not shown) for TI and FTI (DI) that the most probable inter-nuclear distances are around 5 (3) a.u.  between He and middle H, around 7 (5) a.u. between He and left H  and around 3 (3) a.u.  between the two H atoms. Thus, the peak of the KER distribution  for TI and FTI is given roughly by 2/7+ 2/5+1/3.  For DI, where the bound electron is mostly attached on  He  resulting in He$^{+}$, the peak of the KER distribution  is roughly given by 1/5+1/3+1/3.
 
Also, in \fig{fig1} we show that left H$^{+}$ is the faster fragment in all three processes. The slowest fragments are the middle H$^{+}$ for all three processes and  He$^{2+}$ for TI,   He$^{+*}$ for FTI and He$^{+}$ for DI. This is consistent with the two Coulomb repulsive forces on the left H$^{+}$ ion pointing along 180$^{\circ}$ with respect to the +z-axis. The repulsive forces on He$^{2+}$ for TI, He$^{+*}$ for FTI and He$^{+}$ for DI also add up towards 0$^{\circ}$ from the +z-axis. However, the mass of  He compared to H is four times larger. As a result,  He  ends up with a smaller acceleration and hence smaller final kinetic energy compared to the left H$^{+}$. In addition, the repulsive forces on the middle H$^{+}$ from He$^{2+}$ for TI,  He$^{+*}$ for FTI and  He$^{+}$ for DI and from left H$^{+}$  point in opposite directions. As a result, the kinetic energy of the middle H$^{+}$ ion is smaller compared to the left H$^{+}$ ion.

 Moreover we find that the KER distribution of middle H$^{+}$ has a double peak structure for TI and FTI. This double peak is associated with middle H$^{+}$ escaping mainly either along or at an angle with respect to the molecular axis away from He$^{2+}$ for TI and He$^{+*}$ for FTI. The lower (higher) peak in the KER distribution of middle H$^{+}$ corresponds to the middle H$^{+}$ escaping along (at an angle with) the molecular axis. The lower peak is more pronounced for DI, since the force on middle H$^{+}$ from He$^{+}$ in DI is smaller than the force from He$^{2+}$ in TI and He$^{+*}$ in FTI.

 \subsection{Angular distributions}
\begin{figure*}
\includegraphics[scale=0.4]{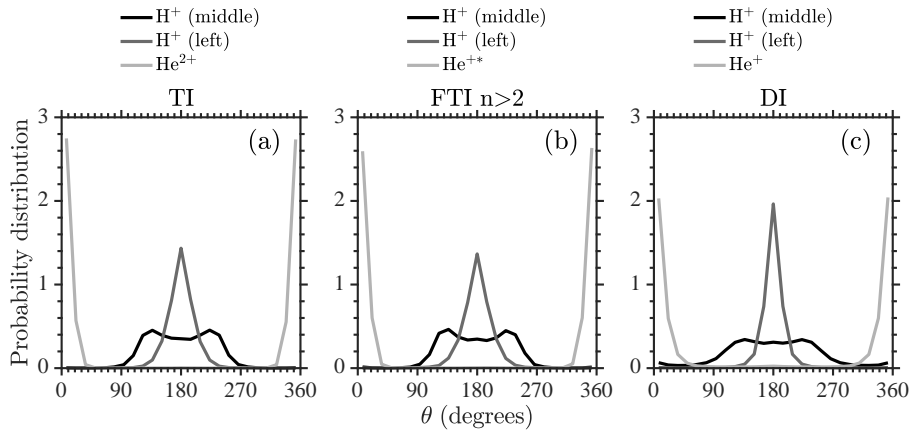}
\centering
\caption{Angular distributions  of the ion fragments produced in (a) triple ionization, (b) ``frustrated" triple ionization and (c) double ionization. The escape along the +z axis corresponds to 0$^{\circ}$. All curves are normalized to one.}
\label{fig2}
\end{figure*}

In \fig{fig2}, we plot the angular distribution of the final  ion fragments for  triple ionization as well as for the most probable channels of ``frustrated" triple ionization and double ionization. For TI and FTI, we find that the angular distribution of middle H$^{+}$ is broader compared to left H$^{+}$. As discussed above, this is consistent with the Coulomb repulsive force on the middle  H$^{+}$ ion being smaller compared to the left H$^{+}$ ion. We expect that another factor contributing to the broader angular distribution of the middle  H$^{+}$ ion   is that the two electrons that tunnel-ionize last mostly move between  He and middle H$^{+}$, before they both (one) escape for TI (FTI). This is due to He having a  higher nuclear charge. The difference between the two angular distributions of the H$^{+}$ ions is even more pronounced for DI. Indeed, the full screening of He$^{2+}$ by the  electron bound in the $n=1$ state results  in He$^{+}$  exerting a force towards 180$^{\circ}$ on the middle H$^{+}$ ion that roughly cancels out the force towards 0$^{\circ}$ from the left H$^{+}$ ion.

\subsection{n quantum number for FTI}
\begin{figure}
\includegraphics[scale=0.3]{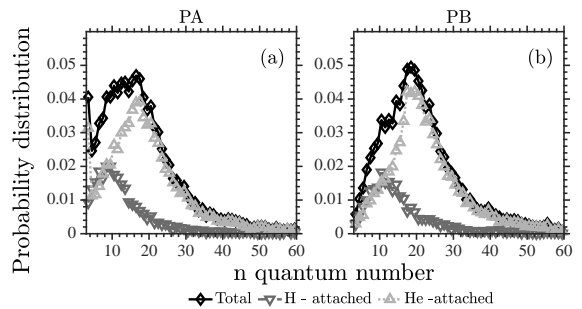}
\centering
\caption{Distribution of the principal n quantum number for pathway A (a) and B (b) of FTI (black solid lines with diamonds). For pathway A and  B of FTI the distribution of the n quantum number is also plotted separately when the electron remains attached to He$^{2+}$ (light grey lines with upwards pointing triangles) and when it remains attached on H$^{+}$ (dark grey lines with downwards pointing triangles).}
\label{fig3}
\end{figure}

Next, we investigate the distribution of the principal n quantum number of the two main pathways of FTI. We find that both pathways A and B with $n>2$ contribute roughly the same to FTI. As already  discussed, we find that formation of a  Rydberg state  is three times more likely for He versus H  attachment. This is shown in \fig{fig3} where we plot the distribution of the principal quantum number n for pathways A and B of FTI. We also find that the distribution of the principal quantum number n peaks around 20 when the electron remains bound in a Rydberg state of He$^{2+}$ versus 10 following attachment on H$^{+}$. This is expected. One assumes  that the electron that tunnel-ionizes last and remains bound in  a Rydberg state  has roughly the same energy for attachment on  He or H. Then, given the dependence of the energy of a hydrogenic atom on the nuclear charge Z and the n number,  it follows that an n number for attachment on H is equivalent to a 2n number for attachment on He. Moreover, we find that the distribution of the n number peaks at higher n values for pathway B versus pathway A. This is consistent with the electron that remains bound in a Rydberg state in pathway B  being the electron that tunnel-ionizes in the initial state. As a result, this electron    upon its return to the nuclei has higher energy compared to the energy that an initially bound electron has when it tunnel-ionizes for the last time and remains  in a Rydberg state in pathway A.

\subsection{Correlation in triple ionization}
Our results suggest a weak effect of the  correlated electron dynamics  in triple ionization during fragmentation of the strongly-driven  triatomic molecule HeH$_{2}^{+}$. Specifically, we find that  for roughly  20\% of triple ionization events  a re-collision takes place where electron 1 transfers energy to both 
initially bound electrons 2 and 3 at the same time. For these events,  the distribution of the maximum of the potential energy between electrons 1 and 2 or electrons 1 and 3 as a function of time  extends up to 2 a.u. (not shown). For the rest of the TI events this maximum of the potential energy 
peaks overwhelmingly around very small values.

\begin{figure}
\includegraphics[scale=0.35]{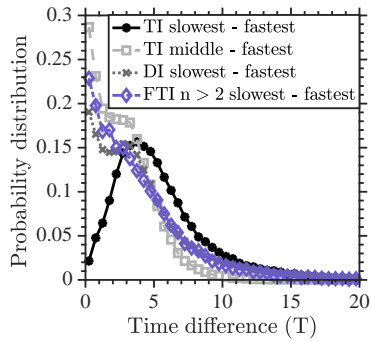}
\centering
\caption{Distribution of the time differences between the  fastest and second fastest electron as well as the fastest and slowest electron in TI and between the fastest and slowest electrons in FTI and DI. }
\label{fig4}
\end{figure}

The weak electronic correlation in TI of strongly-driven HeH$_{2}^{+}$ is also supported by  the distribution of the difference in ionization times of the fastest and second fastest electron as well as the fastest and slowest electron. These two distributions are shown  in \fig{fig4}. We find that the electron that ionizes second has a significant probability to do so with a small time difference from the fastest one. It also has a significant probability to do so with time differences extending from one to four periods of the laser field. In contrast, the last electron to ionize does so with a distribution of  time differences that roughly peaks around  four periods of the laser field.
 This suggests, that the second but mostly the last to ionize electrons escape mainly  due to enhanced ionization and not due to a re-collision. This is expected   for molecules that are fragmenting when driven by long duration and intense pulses. We find this to also be the case for FTI and DI. Indeed, in \fig{fig4} the distribution of the time differences between the two electrons that escape in FTI and DI extends up to  roughly five periods of the laser field, suggesting that FTI and DI take place mostly due to enhanced ionization. 

 \section{Conclusions}
  In this work, we formulate a 3D semi-classical model for a strongly-driven 6-body Coulomb problem to address three-electron dynamics in strongly-driven triatomic molecules.  Since we include the Coulomb singularities, we address how to avoid unphysical autoionization between two bound electrons.  To do so, we develop two criteria that allow us to switch off the Coulomb force between two bound electrons and switch it on when one of the two electrons ionizes or ``quasi-ionizes".  In our current formulation, the two bound electrons  screen each other only indirectly via their interaction with the three nuclei and the other electron. We expect that this is a very good approximation for processes involving electrons in highly excited states before the electrons actually ionize. In this case the electrons screen each other less. Hence, we expect the current formulation to accurately describe mostly TI and FTI
  and less so DI. Indeed, in DI one electron remains bound in the $n=1$ state resulting in higher screening of the nuclear charge during the time that it takes for the last electron to ionize. 
  
   Using this 3D semiclassical model, we address triple and double ionization as well as  ``frustrated" triple ionization in a strongly-driven linear triatomic molecule, namely, HeH$_{2}^{+}$. We find that the electronic correlation in all three ionization processes is weak. Moreover,
  we find that, as for ``frustrated" double ionization, pathways A and B of FTI both contribute to the formation of Rydberg states at the end of the laser field. We also find that the electron that remains bound in FTI is roughly three times more likely to be attached to He$^{2+}$ compared to H$^{+}$. Computing the angular distributions of the final ion fragments in all three ionization processes we find that middle H$^{+}$ escapes with a broader range of angles compared to left H$^{+}$  and He$^{2+}$ for TI, He$^{+*}$ for FTI and He$^{+}$ for DI. This is mainly due to the Coulomb repulsive forces on the middle H$^{+}$ ion from the other two ions pointing in opposite directions. Thus,  the resultant force  has a smaller  magnitude compared to the forces on the other two ions. For DI these opposite pointing Coulomb forces on the middle H$^{+}$ are even more comparable in magnitude. This is due to the screening of He$^{2+}$ from the  $n=1$ bound electron. As a result, the angular distribution of the middle H$^{+}$ ion is even broader for DI compared to TI and FTI.
  
  Our future studies will focus on generalizing our current formulation to include effective potentials that will account for the electronic repulsion between two electrons during the time that they are both bound. This will allow us to also consider processes with two bound electrons, such as ``frustrated" double  ionization.
 
\section{acknowledgment} A. E. acknowledges the EPSRC grant no. N031326.
V. P. M. acknowledges support from SERB grant no.
SRG/2019/001169.
The authors acknowledge the use of the UCL Myriad High Performance Computing Facility (Myriad@UCL), and associated support services, in the completion of this work.



\end{document}